\documentclass{amsproc}[11pt]
\usepackage{graphicx,amssymb}

\theoremstyle{definition}

\theoremstyle{remark}

\numberwithin{equation}{section}

\newcommand{\be}{\begin{equation}}
\newcommand{\ee}{\end{equation}}
\newcommand{\bea}{\begin{eqnarray}}
\newcommand{\eea}{\end{eqnarray}}
\newcommand{\beas}{\begin{eqnarray*}}
\newcommand{\eeas}{\end{eqnarray*}}

\newcommand{\htt}{\hat{t}}
\newcommand{\hp}{\hat{\phi}}

\begin{document}

\title{Local bulk operators in AdS/CFT and the fate of the BTZ singularity}

\author{Alex Hamilton}
\address{Department of Physics, Columbia University, New York NY 10027 USA}
\curraddr{Department of Mathematics and Applied Mathematics,
University of Cape Town, Rondebosch 7700, South Africa}
\email{ahamil@gmail.com}
\thanks{AH was supported by a Columbia University Initiatives in Science and
Engineering grant and by DOE grant DE-FG02-92ER40699}

\author{Daniel Kabat}
\address{Department of Physics, Columbia University, New York NY 10027 USA}
\email{kabat@phys.columbia.edu}
\thanks{DK was supported by DOE grant DE-FG02-92ER40699}

\author{Gilad Lifschytz}
\address{Department of Mathematics and Physics and CCMSC, University of Haifa
at Oranim, Tivon 36006 ISRAEL}
\email{giladl@research.haifa.ac.il}
\thanks{GL is supported in part by Israeli science foundation grant number 568/05}

\author{David A.\ Lowe}
\address{Department of Physics, Brown University, Providence RI 02912 USA}
\email{lowe@brown.edu}
\thanks{DL was supported by DOE grant DE-FG02-91ER40688-Task A}


\dedicatory{Presented by DK at the 2007 Sowers workshop and GL at
the pre-strings 2007 workshop}


\begin{abstract}
This paper has two parts.  First we review the description of local
bulk operators in Lorentzian AdS in terms of non-local operators in
the boundary CFT.  We discuss how bulk locality arises in pure AdS
backgrounds and how it is modified at finite $N$.  Next we present
some new results on BTZ black holes: local operators can be defined
inside the horizon of a finite $N$ BTZ black hole, in a way that
suggests the BTZ geometry describes an average over black hole
microstates, but with finite $N$ effects resolving the singularity.
\end{abstract}

\maketitle

\section{Introduction\label{Intro}}

Quantum gravity in asymptotically anti-de Sitter space is dual to a
conformal field theory on the boundary of AdS \cite{Maldacena:1997re}.
One of the most interesting questions raised by this duality is: how
does approximately local bulk gravitational physics emerge from the
CFT?

To address this one needs some way of probing local physics in the
bulk.  We are mostly interested in the semiclassical limit of small
Planck length.  In this limit we should be able to recover the
traditional results of quantum field theory in curved space
\cite{Birrell:1982ix}.  So it's natural to ask: how can a local
quantum field in the bulk of AdS be represented in the boundary CFT?

This question was addressed in
\cite{Banks:1998dd,Balasubramanian:1999ri,Bena} and was further
developed by the present authors in
\cite{Hamilton:2006az,Hamilton:2005ju,Hamilton:2006fh}.  In the latter
works it was shown that local operators in the bulk could be
represented as non-local operators in the CFT.  The CFT operators turn
out to have support on a compact region of the complexified boundary.
This representation makes several properties manifest.  It makes it
clear why bulk locality is exact at large $N$, but
breaks down at finite $N$, in exactly the manner required by
holography.  It also provides a simple CFT description of the horizon
and singularity of a BTZ black hole in the large $N$ limit.

An outline of this paper is as follows.  In section \ref{construction}
we review the representation of local bulk operators in terms of
operators on the complexified boundary.  In section \ref{locality} we
use these boundary operators to discuss bulk locality and
holography from the point of view of the CFT.  In section \ref{BTZ} we
extend the construction to the BTZ black hole and discuss the
horizon and singularity in the large $N$ limit.  We conclude in
section \ref{BTZ2} with some speculation on the fate of the horizon
and singularity at finite $N$.  Sections \ref{construction} -- \ref{BTZ}
are a review; the results in section \ref{BTZ2} are new.

\section{Local operators in the semiclassical limit\label{construction}}

In Poincar\'e coordinates the metric on Lorentzian AdS${}_D$ is
\be
\label{AdSmetric}
ds^2 = {R^2 \over Z^2} \left(-dT^2 + \vert d{\bf X} \vert^2 + dZ^2\right)\,.
\ee
Here $R$ is the AdS radius.  The Poincar\'e horizon is at $Z =
\infty$, while the CFT${}_{d = D-1}$ lives on the boundary at $Z = 0$.
Consider a scalar field of mass $m$ in AdS, with normalizable
fall-off near the boundary.
\[
\phi(T,{\bf X},Z) \sim Z^\Delta \phi_0(T,{\bf X}) \quad \hbox{\rm as $Z \rightarrow 0$}
\]
The parameter $\Delta$ is related to the mass of the field by
\[
\Delta = {d \over 2} + \sqrt{{d^2 \over 4} + m^2 R^2}\,.
\]
We will refer to $\phi_0$ as the boundary field.  It's dual to an
operator of dimension $\Delta$ in the CFT.
\be
\label{duality}
\phi_0(T,{\bf X})_{\rm SUGRA} \leftrightarrow {\mathcal O}(T,{\bf X})_{\rm CFT}
\ee
The question is, can we express $\phi$ in terms of $\phi_0$?  If so,
then we can use (\ref{duality}) to find the CFT operator dual to a
local operator in the bulk.

For now we'll study this in the semiclassical limit
\beas
&& \ell_S,\ell_P \rightarrow 0 \qquad \hbox{\rm in the bulk} \\
&& N, \lambda \rightarrow \infty \qquad \hbox{\rm on the boundary}
\eeas
Here $\ell_S$ and $\ell_P$ are the bulk string and Planck lengths,
while $N$ and $\lambda$ are parameters for some kind of 't Hooft
large-$N$ expansion in the CFT whose details won't matter for us.
The basic idea is to represent
\[
\phi(T,{\bf X},Z) = \int dT' d^{d-1}X' \, K(T',{\bf X}' \vert T,{\bf X},Z) \, \phi_0(T',{\bf X}')
\]
using a kernel or smearing function $K$.  Since AdS has a timelike
boundary, this is not a standard Cauchy problem, and neither existence
nor uniqueness of $K$ is guaranteed.  Indeed in \cite{Hamilton:2006az}
we discuss examples where both existence and uniqueness are violated.

A cure for these problems, at least in a pure semiclassical AdS
background, is to make a Wick rotation to de Sitter space.  Define a
new set of boundary spatial coordinates by setting ${\bf X} = i {\bf
Y}$.  This turns the AdS metric (\ref{AdSmetric}) into
\be
\label{dSmetric}
ds^2 = {R^2 \over Z^2} \left(-dT^2 - \vert d{\bf Y} \vert^2 + dZ^2\right)\,.
\ee
This is de Sitter space in flat FRW coordinates, with $Z$ playing the
role of conformal time.  The past boundary of de Sitter space is at $Z
= 0$.

\begin{figure}
\begin{center}
\includegraphics[width=4in]{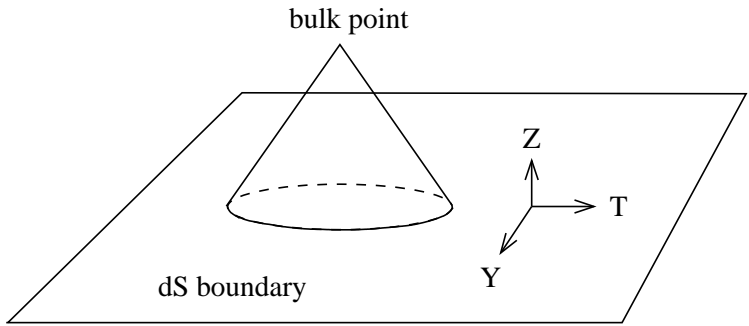}
\end{center}
\caption{\label{fig:cone}The field at a bulk point in de Sitter space can be
expressed in terms of data on the past de Sitter boundary.  The slice $Y = 0$
also describes a region in AdS.  So we can also regard this as expressing the
field in AdS in terms of data on the complexified AdS boundary.}
\end{figure}

In de Sitter space we have a standard Cauchy problem.  As shown in
Fig.~\ref{fig:cone} we can use a retarded de Sitter Green's function
to solve for the bulk field in terms of data on the past boundary.
The explicit analytic expressions are pretty simple: the field at a point
in AdS can be expressed as \cite{Hamilton:2006fh}
\bea
\nonumber
\phi(T,{\bf X},Z) & = & \frac{\Gamma\left(\Delta - \frac{d}{2} + 1 \right)}
{\pi^{d/2} \ \Gamma(\Delta - d + 1)}
\int_{T'{}^2 + \vert {\bf Y}'\vert^2 < Z^2}\!\!\!\! dT' d^{d-1}Y' \,
\left({Z^2 - T'{}^2 - \vert {\bf Y}' \vert^ 2 \over Z}\right)^{\Delta - d} \\*[5pt]
\label{kernel}
& & \qquad \qquad \phi_0(T + T', {\bf X} + i{\bf Y}')\,.
\eea
Note that we have to integrate over a compact region of the de Sitter
boundary (the region inside the past light-cone of the bulk point).
Equivalently we integrate over a compact region of the complexified
AdS boundary (the region spacelike separated from the bulk point).

\begin{figure}
\begin{center}
\includegraphics[width=5in]{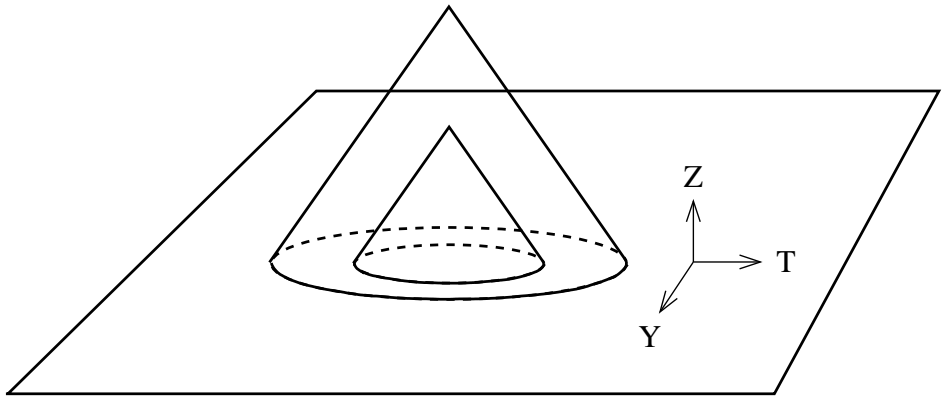}
\end{center}
\caption{\label{fig:cone2}Smearing functions for two bulk points
separated only in the $Z$ direction.  The smearing functions overlap
on the boundary, nonetheless the smeared operators commute at
infinite $N$.}
\end{figure}

By construction this lets us reproduce bulk correlation functions in
the semiclassical limit.
\be
\label{convolve}
\langle \phi(x_1,Z_1) \phi(x_2,Z_2) \rangle_{SUGRA} = \int dx_1' dx_2' \, K(x_1' \vert x_1,Z_1)
K(x_2' \vert x_2,Z_2) \langle {\mathcal O}(x_1') {\mathcal O}(x_2') \rangle_{CFT}
\ee
This is guaranteed to work, just because $\phi_0$ and ${\mathcal O}$
have identical correlators.  Although somewhat trivial, this result
has an interesting corollary.  In the semiclassical limit of vanishing
Planck length, bulk operators will commute at spacelike separation.
Therefore the corresponding smeared boundary operators will commute
with each other.  This is true even though, as shown in Fig.~\ref{fig:cone2},
the smearing functions might overlap on the boundary.

\section{Bulk locality and holography at finite $N$\label{locality}}

What we've done so far is exact in the semiclassical limit; it can be
regarded as a set of statements about free wave equations in a pure
AdS background.  In this section we'll remain in a pure AdS
background, but ask what happens at finite $N$.

First we need to decide what smearing functions to use.
One possibility is to use the same smearing functions at
finite $N$.  For example, in ${\mathcal N} = 4$ Yang-Mills the
operator
\be
\label{finiteN}
\Phi(T,{\bf X},Z) = \int dT'd^3X' \, K(T',{\bf X}' \vert T,{\bf X},Z) \, {\rm Tr} F^2
\ee
can be defined at any $N$, where $K$ is the kernel appearing in
(\ref{kernel}).  In a pure AdS background we believe these must be the
right operators to use at finite $N$, just because the construction is
singled out by the symmetries.  To see this, introduce a distance
function on AdS.
\be
\label{DistanceFunction}
\sigma(T,{\bf X},Z \vert T',{\bf X}',Z') = \cosh \left({\hbox{\small geodesic distance} \over R}\right)
= {Z^2 + Z'{}^2 + \vert {\bf X} - {\bf X}' \vert^2 - (T - T')^2 \over 2 Z Z'}
\ee
$\sigma$ is invariant under AdS isometries.  In Poincar\'e coordinates
the isometry
\[
(T,{\bf X},Z) \rightarrow \lambda (T,{\bf X},Z)
\]
acts as a scale transformation on the boundary.  $Z$ has conformal
weight $-1$, and since $K \sim \lim_{Z' \rightarrow 0} (\sigma
Z')^{\Delta - d}$, we see that $K$ transforms covariantly under AdS
isometries with conformal weight $d - \Delta$.  But given an operator
of dimensions $\Delta$, this is exactly what we need for $\int d^dx \,
K {\mathcal O}$ to behave like a scalar field in the bulk!  That is,
{\em the smearing functions we have defined provide the unique
covariant way to map a primary field in the CFT to a scalar field in
the bulk.}\footnote{Thus the smearing functions should have a purely
group-theoretic interpretation in terms of representations of the
(complexified) isometry group $SO(d,2)$, along the lines of
\cite{VK}.  We are grateful to Djordje Minic for discussions on this
point.}\footnote{ To avoid a possible confusion: by construction,
the operators defined in (\ref{finiteN}) satisfy a free wave
equation.  So they do not have the right interactions to be
identified with the bulk dilaton field at finite $N$.  Nevertheless,
they should serve as good probes of local physics in the bulk.}

This might seem very strange from the point of view of holography.  In
(\ref{finiteN}) we've defined a continuous infinity of bulk operators.
How can this be compatible with the holographic bound
\cite{Susskind:1998dq}, which should only allow a finite number of
degrees of freedom in any given region in the bulk?

\begin{figure}
\begin{center}
\includegraphics[width=2in]{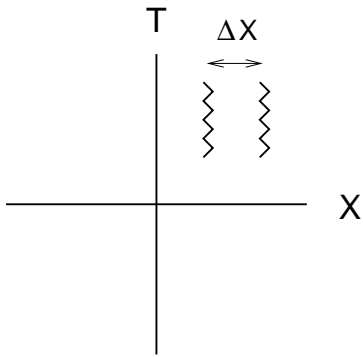}
\end{center}
\caption{\label{fig:cuts}The smearing functions have support on the
two jagged lines.  For $\vert \Delta X \vert > Z$ they are spacelike
separated.}
\end{figure}

We believe the resolution is that at finite $N$ not all the
operators defined in (\ref{finiteN}) commute at spacelike
separation.  To see this, consider a fixed-$T$ hypersurface in the
bulk, with operators placed at some radial position $Z$.  The smearing
functions we have defined have an extent in the time direction
$\Delta T = Z$.\footnote{They also have an extent in the imaginary
spatial directions, but as shown in \cite{Hamilton:2006fh} it's only
the extent in time that matters here.}  As shown in
Fig.~\ref{fig:cuts} the operators will be spacelike separated on the
boundary provided the bulk operators have a spatial separation
$\vert \Delta X \vert > Z$.
In this case the operators are guaranteed to commute,
just by locality of the boundary theory.  At finite $N$ we do not
expect operators separated by $\vert \Delta X \vert < Z$ to commute.
(This is unlike the semiclassical situation discussed in section
\ref{construction}, where operators could commute even though they
overlapped.)  So we expect $1/Z^{d-1}$ commuting operators per
coordinate area on the boundary.  Using the AdS metric
(\ref{AdSmetric}), this means we expect $1/R^{d-1}$ commuting
operators per proper area in the bulk.  Equivalently we expect one
commuting operator per unit cell, where the cell volume $\sim
R^{d-1}$.  At finite $N$, bulk locality breaks down on distances set
by the AdS radius of curvature!

This is a bit disturbing, in that it seems the commuting operators we
can build from ${\rm Tr} F^2$ aren't sufficient to describe a local
bulk dilaton on distances less than an AdS radius.  We don't have a
complete resolution of this puzzle.  It's probably too strong a
condition to require that all operators describing the dilaton commute
exactly.  Take the set of commuting operators built from ${\rm Tr}
F^2$.  Perhaps there are additional operators which can be used to
describe the dilaton, which do not all commute, but whose commutators
are so small at low energies that they can be ignored.\footnote{More
precisely, we expect matrix elements of the commutators between
low-energy states to be small.}  The operators built from ${\rm Tr}
F^2$ using our smearing functions are good first candidates for the
job, since their commutators do vanish in the large-$N$ limit.  But it
could also be that operator mixing is important, so that operators besides
${\rm Tr} F^2$ can contribute.

One might also wonder about the holographic bound.  Here things work
out very nicely.  The holographic bound in AdS states that the entropy
per longitudinal coordinate area is bounded by $s < N^2/Z^{d-1}$.
Each operator in the CFT gives an entropy density $\sim 1/Z^{d-1}$, so
bound is saturated if we have $N^2$ commuting operators in the CFT.
This seems quite reasonable, as ${\mathcal N} = 4$ Yang-Mills involves
$N \times N$ matrices and has a central charge $\sim N^2$.

\section{Semiclassical BTZ black hole\label{BTZ}}

We now turn to excitations of AdS${}_3$, in particular we will
study non-extremal BTZ black holes.  But first let's consider AdS${}_3$
in accelerating or Rindler-like coordinates.
\bea
\label{AdS3}
&& ds^2 = - {r^2 - r_0^2 \over R^2} dt^2 + {R^2 \over r^2 - r_0^2} dr^2 + r^2 d\phi^2 \\
\nonumber
&& -\infty < t,\phi < \infty \qquad\quad 0 < r < \infty
\eea
Here $r_0$ is an arbitrary parameter with units of length.\footnote{We
denoted this parameter $r_+$ in our previous work.}  We'll frequently
work in terms of the rescaled coordinates
\[
\htt = r_0 t / R^2 \qquad \qquad \hp = r_0 \phi / R\,.
\]
It's straightforward to construct smearing functions in these coordinates.
The Wick rotation $\phi = i y$ turns the AdS${}_3$ metric (\ref{AdS3})
into de Sitter space, now expressed in static coordinates.
\bea
\label{dS3}
&& ds^2 = - {r^2 - r_0^2 \over R^2} dt^2 + {R^2 \over r^2 - r_0^2} dr^2 - r^2 dy^2 \\
\nonumber
&& -\infty < t < \infty \qquad y \approx y + 2\pi R/r_0 \qquad 0 < r < \infty
\eea
(The periodicity in $y$ is necessary to avoid a singularity at $r = 0$.)
One can use a retarded de Sitter Green's function to construct
smearing functions in AdS${}_3$.  Alternatively, one can just
translate our previous result (\ref{kernel}) into Rindler coordinates,
to find\footnote{The boundary field $\phi_0$ is defined slightly differently
in Rindler coordinates: $\phi(t,r,\phi) \sim \phi_0(t,\phi) / r^\Delta$
as $r \rightarrow \infty$.}
\bea
\label{RindlerSmear}
&& \phi(t,r,\phi) = \int dt'dy' \, K_{\rm Rindler}(t+t',\phi+iy' \vert t,r,\phi) \, \phi_0(t+t',\phi+iy') \\
\nonumber
&& K_{\rm Rindler} = {(\Delta - 1) 2^{\Delta - 2} \over \pi R^3}
\lim_{r' \rightarrow \infty} \left({\sigma \over r'}\right)^{\Delta - 2}
\eea
where $\sigma$ is the invariant distance between $(t,r,\phi)$ and $(t+t',r',\phi+iy')$, and the integral is over spacelike separated points
on the complexified AdS boundary.
A BTZ black hole can be obtained from AdS${}_3$ by identifying $\phi \approx
\phi + 2\pi$.  This produces an orbifold singularity at $r = 0$.  But making
this identification doesn't change the smearing functions at all: if
the boundary field has the necessary periodicity, so will the bulk field.
So we can use the same smearing functions (\ref{RindlerSmear}) in a BTZ
background.

In the semiclassical limit this gives a clear picture of the BTZ
horizon and singularity.  First, the horizon.  The integral in
(\ref{RindlerSmear}) is over points on the complexified boundary that
are spacelike separated from the bulk point.  As the bulk point
approaches the (future, past) horizon the integration region extends
to ($t = +\infty$, $t = -\infty$).  Thus to probe the horizon requires
an infinite time from the boundary point of view.  This fits nicely
with the bulk definition of a horizon, as bounding the region where
light rays cannot escape to infinity.\footnote{One can also describe
bulk points that are located inside the horizon
\cite{Hamilton:2006fh}.  However this requires the use of smearing
functions with support on both the left and right boundaries of the
extended Kruskal diagram.  For a bulk point inside the future
horizon the smearing function extends to $t = + \infty$ on the right
boundary and $t = -\infty$ on the left boundary, where time has the
same orientation on the left and right.  From the boundary point of
view this means we are using operators that act on both copies of
the thermofield-doubled CFT.}

What about the BTZ singularity?  With scalar fields as probes we cannot
directly study the bulk geometry.  However it turns out that the orbifold
singularity generates a divergence in scalar correlators.  To see this,
we use the fact that in the semiclassical limit we can make the bulk
correlator periodic with an image sum \cite{Lifschytz:1993eb}.
\be
\label{BTZimage}
\langle \phi(t,r,\phi) \phi(t',r',\phi') \rangle_{\rm BTZ} = \sum_{n=-\infty}^\infty \langle \phi(t,r,\phi)
\phi(t',r',\phi' + 2 \pi n)\rangle_{\rm AdS}
\ee
But $r = 0$ is a fixed point of the isometry $\phi \rightarrow \phi +
{\rm const.}$, so correlators in AdS are $\phi$-independent at $r =
0$.  If we compute a bulk correlator in a BTZ background, the image
sum diverges when one of the bulk points is located at the
singularity.

The same divergence arises from the boundary point of view.  The CFT
dual to AdS${}_3$ in Rindler coordinates lives on ${\mathbb R}^{1,1}$,
while the CFT dual to BTZ lives on ${\mathbb R} \times S^1$.  In the
semiclassical limit the BTZ boundary correlator can be given the
necessary periodicity with an image sum.
\[
\langle \phi_0(t,\phi) \phi_0(t',\phi') \rangle_{\rm BTZ} = \sum_{n=-\infty}^\infty \langle \phi_0(t,\phi)
\phi_0(t',\phi' + 2 \pi n)\rangle_{\rm AdS}
\]
To recover a bulk correlator we convolve this with our smearing
functions as in (\ref{convolve}).  Again the image sum diverges when
one of the bulk points is located at the singularity.  So we also get a
divergent correlator at $r = 0$ from the boundary point of view.

For future reference it's useful to study the divergence in a little
more detail.  Consider a point in AdS located near $r = 0$, at
\[
(t,r,\phi) \quad \hbox{\rm with $t = 0$, $r \rightarrow 0$, $\phi = 0$}
\]
and a second point in AdS located near the boundary, at
\[
(t',r',\phi') \quad \hbox{\rm with $r' \rightarrow \infty$.}
\]
The invariant distance (\ref{DistanceFunction}) between these points
is approximately
\[
\sigma \approx {r' \over r_0} \left({r \over r_0} \cosh \hp' + \sinh \htt'\right)\,.
\]
The AdS bulk correlation function \cite{Ichinose:1994rg} decays
exponentially at large $\phi'$.
\bea
\nonumber
\langle \phi(t,r,\phi) \phi(t',r',\phi') \rangle_{\rm AdS}
& = & {1 \over 4 \pi R \sqrt{\sigma^2 - 1}}
{1 \over \left(\sigma + \sqrt{\sigma^2 - 1}\right)^{\Delta - 1}} \\
\nonumber
& \approx & {(r_0 / 2 r')^\Delta \over 2 \pi R \left({r \over r_0} \cosh \hp' + \sinh \htt' \right)^\Delta} \\
\label{LargePhi}
& \sim & e^{-\Delta \vert\hp'\vert} \quad \hbox{for \hspace{1mm} $\vert\hp'\vert > \hp_{\rm max} \sim \log (r_0/r)$}
\eea
This behavior means the BTZ image sum (\ref{BTZimage}) is cut off at
$\vert n \vert \sim {R \over 2 \pi r_0} \log {r_0 \over r}$, which in turn means
the BTZ correlator diverges logarithmically as $r \rightarrow 0$.\footnote{This
corrects a normalization error in \cite{Hamilton:2006fh}.}
\be
\label{BTZdivergence}
\langle \phi(t,r,\phi) \phi(t',r',\phi') \rangle_{\rm BTZ} \sim {\log(r_0/r) \over 2 \pi^2 r_0}
\left({r_0 \over 2 r' \sinh \htt'}\right)^\Delta \quad \hbox{\rm as $r \rightarrow 0$}
\ee

\section{BTZ at finite $N$\label{BTZ2}}

What happens to the BTZ black hole at finite $N$?  Although definitive
statements are hard to come by, there are a few interesting
observations to make.  First, there's the issue of what smearing
functions to use.  In a pure AdS background we gave a symmetry
argument that the same smearing functions should be used at any value
of $N$.  For excited states such as BTZ, where the symmetries are
broken, this argument is not valid.\footnote{From the boundary point
of view BTZ is related to a CFT on a Euclidean 2-torus, and the
bulk-to-boundary map could depend on the modular parameter of the
torus in a way that cannot be determined from modular invariance.}
Nonetheless one might be tempted to use the same semiclassical
smearing functions for BTZ even at finite $N$.  This is indeed a
reasonable prescription for bulk points outside the horizon.  However
as the bulk point approaches the (future) horizon the semiclassical
smearing functions extend to $t = + \infty$ on the boundary, and for
points inside the horizon they grow exponentially with
time.\footnote{This is the behavior on the right boundary.  For points
inside the future horizon the smearing function also extends to $t =
- \infty$ on the left boundary, and grows as $e^{-(\Delta - d) \htt}$
in the far past.}
\[
K(\htt,\hp \vert \cdot) \sim e^{(\Delta - d)\htt} \quad \hbox{\rm as $t \rightarrow \infty$}
\]
In the semiclassical limit this causes no problems.  CFT correlators
at infinite $N$ decay exponentially \cite{Birmingham:2001pj}, making
the convolution of a boundary correlator with a smearing function
well-defined.
\[
\langle \phi_0(\htt,\hp) \phi_0(0,0) \rangle_{BTZ} \sim e^{-\Delta \htt}
\]
But when $N$ is finite such behavior cannot persist indefinitely.  At
finite $N$, the CFT on ${\mathbb R} \times S^1$ has a finite thermal
entropy.  Eventually the discrete spectrum of the CFT becomes
important and causes correlators to oscillate quasi-periodically
rather than decay exponentially \cite{Dyson:2002nt}.  So we could not
hope to use our semiclassical smearing functions to reproduce sensible
bulk correlators.  It seems there are a couple options.
\begin{enumerate}
\item Perhaps this supports the fuzzball picture of Mathur and
collaborators \cite{Mathur:2005zp}, in which the geometry inside the
horizon differs radically from what one would expect based on the
traditional black hole metric.  In this case one would need to know
the exact microstate of the black hole to make sense of the interior
of the horizon.
\item Maybe there is some modification to the smearing functions which
gives sensible bulk correlators even for points inside the horizon.
``Sensible'' means with small $1/N$ corrections to the
semiclassical result, except near $r=0$, where the divergence
(\ref{BTZdivergence}) should be smoothed out.  This would support
the picture that the semiclassical geometry is a good description,
even inside the horizon, but with quantum gravity effects resolving
the singularity.
\item It could be that options 1 and 2 are compatible, if
one is able to recover the semiclassical BTZ metric from the
fuzzball picture by a suitable averaging procedure
\cite{Alday:2006nd,Balasubramanian:2007zt}.
\end{enumerate}
We conclude by presenting a prototype construction to show how option
3 might be realized.  The difficulty is that for points inside the
horizon the semiclassical smearing functions grow exponentially with
time.  A simple cure is to define a modified smearing function
$\widetilde{K}$ which vanishes if $\htt$ is larger than some cutoff
time $\htt_{\rm max}$:
\[
\widetilde{K}(\htt,\hp \vert \cdot) =
\left\lbrace\begin{array}{ll}
K(\htt,\hp \vert \cdot) & \quad\hbox{\rm if $\vert \htt \vert < \htt_{\rm max}$} \\
0 & \quad\hbox{\rm otherwise}
\end{array}\right.
\]
(We impose the cutoff on both the left and right boundaries of the
black hole.)  We can use this modified smearing function to define --
purely within the CFT -- a set of ``bulk operators.''
\be
\label{ModOperators}
\widetilde{\phi}(t,r,\phi) \stackrel{\mathrm{def}}{=} \int dt'd\phi' \, \widetilde{K}(t',\phi'
\vert t,r,\phi) \,
{\mathcal O}(t',\phi')
\ee
With the cutoff in place, the operators $\widetilde{\phi}$ have finite
correlation functions.

Imposing a cutoff in this way might seem very arbitrary.  But in fact
there is a good physical motivation for the cutoff which sets an upper
bound on $t_{\rm max}$.  With
\[
n(E) = \left(\hbox{\# CFT states with energy $<E$}\right) = e^{S(E)}
\]
the density of states
\[
{dn \over dE} = \beta e^S
\]
implies a spacing between energy levels
\be
\label{spacing}
\Delta E = {1 \over \beta e^S}\,.
\ee
This spacing corresponds to a time (the Heisenberg time
\cite{Srednicki})
\[
t_H = \beta e^S\,.
\]
By this time CFT correlators begin to oscillate quasi-periodically
rather than decay exponentially \cite{Barbon:2003aq,Barbon:2004ce}.
So imposing a cutoff on the smearing functions at $t_{\rm max}
\lesssim t_H$ is the minimal change necessary to obtain well-defined
bulk correlators.\footnote{In general $t_{\rm max}$ should be set by
the time at which CFT correlators begin to behave quasi-periodically.
This could occur before $t_H$, so really $t_H$ is an upper bound on
$t_{\rm max}$.  The main point is that at finite $N$ the upper bound
is finite.  We are grateful to Hong Liu and Massimo Porrati for
discussions on this topic.}  Moreover, such a cutoff has a nice
physical interpretation.  From the CFT point of view measurements with
a duration exceeding $t_H$ can resolve individual microstates of the
CFT.  So putting a cutoff at $t_{\rm max} \lesssim t_H$ implies an
average over microstates.  This means the smearing functions we
constructed based on the classical BTZ geometry break down at the
horizon when $N$ is finite, unless one averages over microstates.
This suggests that the region inside the horizon of the classical BTZ
geometry isn't a good description of any individual microstate of the
black hole.  Rather the region inside the horizon only provides a good
description of ensemble averages over black hole microstates.

With a cutoff at $t_{\rm max} \lesssim t_H$, $\widetilde{K}$
represents the minimal modification to the smearing functions
necessary to plausibly represent BTZ correlators at finite $N$.  In
the semiclassical limit we expect $t_{\rm max} \rightarrow \infty$, so
the bulk operators we have defined have the correct semiclassical
limit.

Ideally at this point we would compute correlation functions of the
operators $\widetilde{\phi}$ in the CFT at finite $N$.  Such a
calculation might be within reach \cite{Witten:2007kt}.  But for the
time being we will regard $t_{\rm max}$ as a fixed {\em ad hoc} cutoff
and study correlators of the operators $\widetilde{\phi}$ in the
large-$N$ limit of the CFT.  Up to small $1/N$ corrections, this
should be a good guide to behavior at finite $N$.

\begin{figure}
\begin{center}
\includegraphics[width=3in]{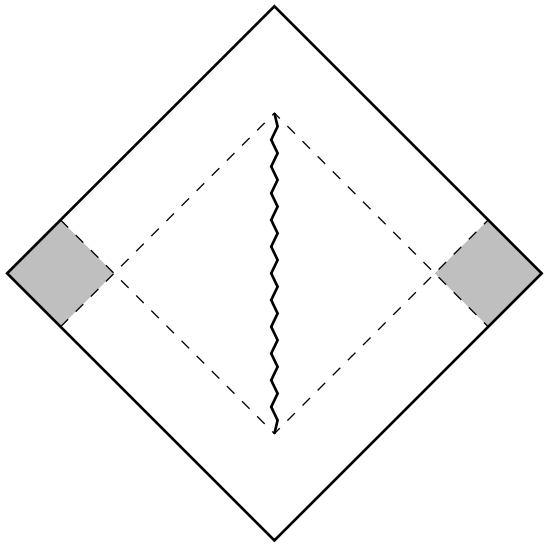}
\end{center}
\caption{\label{fig:Penrose}Penrose diagram of the $(t,\phi)$ plane.  The
support of $\widetilde{K}$ is indicated by the jagged line.  Points in the
shaded region are spacelike separated from the support of
$\widetilde{K}$.  When the smearing function extends from
$-\htt_{\rm max}$ to $+\htt_{\rm max}$ the shaded region is
characterized by $\vert \hp \vert > \htt_{\rm max} + \vert \htt
\vert$.}
\end{figure}

We first work in Rindler coordinates on AdS${}_3$ and consider the
correlation function
\be
\label{AdS3ModCor}
\langle \, \widetilde{\phi}(t,r,\phi) \, \widetilde{\phi}(t',r',\phi') \, \rangle_{\rm AdS}
\ee
between a bulk operator located at $(t=0,r,\phi=0)$ and an operator
near the boundary at $(t',r',\phi')$ with $r' \rightarrow \infty$. In
the semiclassical limit the boundary correlator
\be
\label{BdyCorrelator}
\langle \phi_0(\htt,\hp) \phi_0(0,0) \rangle_{\rm AdS} = {(r_0^2/2)^\Delta \over
2\pi R \left(\cosh \hp - \cosh \htt\right)^\Delta}
\ee
decays exponentially at spacelike separation.  So the bulk-boundary
correlator (\ref{AdS3ModCor}) will be exponentially small provided the
boundary point is spacelike separated from the support of the smearing
function.  As shown in Fig.~\ref{fig:Penrose}, with the modified smearing
functions the cutoff at $t_{\rm max}$ means that -- even for a bulk point
inside the Rindler horizon -- the correlator will decay exponentially at
large $\phi'$.

To extend this discussion to BTZ we use the image charge construction
(\ref{BTZimage}).  For points inside the horizon $\htt_{\rm max}$
serves to cut off the AdS correlator at $\vert \hp' \vert \approx
\htt_{\rm max}$.  However as shown in (\ref{LargePhi}) the AdS
correlator is already exponentially small when $\vert \hp' \vert >
\hp_{\rm max}$.  Thus when we perform the image sum there are two
possible regimes.
\begin{enumerate}
\item For $\htt_{\rm max} > \hp_{\rm max}$, or equivalently for
$r > r_0 e^{- \htt_{\rm max}}$,
the additional cutoff at $\htt_{\rm max}$ isn't important.  So using
$\widetilde{K}$ rather than $K$ makes a negligible change to the BTZ
correlator away from $r = 0$.  We can therefore probe a large region
inside the BTZ horizon, roughly the region
\[
r_0 e^{- \htt_{\rm max}} < r < r_0\,,
\]
using only a finite time interval on the boundary, and without seeing
significant deviations from the semiclassical result.  Strictly speaking
this means there is no horizon, at least not in the sense of section
\ref{BTZ} where the horizon corresponded to integration over infinite time.
\item However for $\htt_{\rm max} < \hp_{\rm max}$, or equivalently for
\[
0 < r < r_0 e^{- \htt_{\rm max}}\,,
\]
the additional cutoff at $\htt_{\rm max}$ is crucial.  It serves to
regulate the image sum, cutting it off at $\vert n \vert \approx
{R \over 2\pi r_0} \htt_{\rm max}$ so that
\[
\langle \, \widetilde{\phi}(t,r,\phi) \, \widetilde{\phi}(t',r',\phi') \, \rangle_{\rm BTZ}
\approx {\htt_{\rm max} \over 2 \pi^2 r_0} \left({r_0 \over 2 r' \sinh \htt'}\right)^\Delta\,.
\]
Note that the correlator is independent of $r$, and the divergence at
$r = 0$ has been eliminated!
\end{enumerate}
We find it appealing that the same effect that eliminates the horizon
also gets rid of the divergence.  Note that the effects we have
discussed are very robust: they are independent of any details of the
CFT and only rely on the generic thermal behavior (\ref{spacing}).
Our results are compatible with the detailed study of extremal BTZ
black holes in \cite{Balasubramanian:2007qv}.

To summarize: we've defined a set of operators in the CFT
(\ref{ModOperators}) which should have well-defined correlation
functions even at finite $N$.  As a guide to the behavior of these
operators we studied their correlation functions in the large $N$
limit.  For bulk points well outside the horizon the cutoff at $t_{\rm
max}$ has no effect on the smearing functions.  For bulk points
inside the horizon but well away from the singularity the cutoff at
$t_{\rm max}$ makes an exponentially small change in correlators.  But
for points very near $r=0$ the cutoff at $t_{\rm max}$ becomes
important and renders correlation functions finite.  It seems
reasonable that working with the true finite-$N$ correlation functions
of the CFT, rather than their semiclassical large-$N$ limit, should
only make a small change in these results.  If so, this would support
option 3: after a suitable average over microstates, enforced by a
cutoff at $t_{\rm max} \lesssim t_H$, the semiclassical BTZ geometry
becomes a good description, even inside the horizon, but with quantum
gravity effects resolving the singularity.

\bigskip
\goodbreak
\centerline{\bf Acknowledgements}
\noindent
DK is grateful to Mark Sowers and the organizers of the 2007 Sowers
workshop for a delightful and stimulating conference.  GL would like
to thank Vijay Balasubramanian and especially Masaki Shigemori for
discussions and the organizers of pre-strings 2007 for a stimulating
conference.


\newcommand{\etalchar}[1]{$^{#1}$}
\providecommand{\bysame}{\leavevmode\hbox to3em{\hrulefill}\thinspace}
\providecommand{\MR}{\relax\ifhmode\unskip\space\fi MR }
\providecommand{\MRhref}[2]{%
  \href{http://www.ams.org/mathscinet-getitem?mr=#1}{#2}
}
\providecommand{\href}[2]{#2}

\end{document}